\documentclass{ws-ijmpa}

\def\D0bar{\overline D{}^0}

\def\beq{\begin{equation}}
\def\eeq{\end{equation}}
\def\bea{\begin{eqnarray}}
\def\eea{\end{eqnarray}}

\newcommand{\Dz}{D^0}
\newcommand{\Dzb}{\overline{D^0}}
\newcommand{\DzDzb}{D^0-\overline{D^0}}
\newcommand{\BzBzb}{B^0-\overline{B^0}}

\begin{document}

\markboth{Alexey A Petrov}
{CHARM MIXING IN THE STANDARD MODEL AND BEYOND}

%
\catchline{}{}{}{}{}
%

\title{
CHARM MIXING IN THE STANDARD MODEL AND BEYOND
}

\author{ALEXEY A PETROV}

\address{Department of Physics and Astronomy, Wayne State University\\
Detroit, MI 48201, USA\\
apetrov@wayne.edu}

\maketitle


\begin{abstract}
The motivation most often cited in searches for $\DzDzb$ mixing and CP-violation in 
charm system lies with the possibility of observing a signal from New Physics which 
dominates that from the Standard Model. We review recent theoretical predictions and
experimental constraints on $\DzDzb$ mixing parameters, concentrating on possible
effects of New Physics.

\end{abstract}

\ccode{PACS numbers: 11.25.Hf, 123.1K}

\section{Introduction}

Quantum mechanical meson-antimeson oscillations are sensitive to heavy degrees 
of freedom which propagate in the underlying mixing amplitudes. Comparing observed 
meson mixing with predictions of the Standard Model (SM), modern experimental studies
would be able to constrain models of New Physics (NP). Yet, extensive precission data 
from B-factories and the Tevatron collider show that the large SM mixing succesfully 
describes all available experimental data in $B_d$ and $B_s$ oscillations. The only 
flavor oscillation not yet observed is that of the charmed meson $D^0$, where SM 
mixing is very small and the NP component can stand out. This situation is an exact 
opposite to what happens in the $B$ system, where $\BzBzb$ mixing measurements are 
used to constrain top quark couplings.

Together with the one loop Standard Model effects\cite{Petrov:1997ch} 
NP effects can contribute to $\Delta C = 1$ (decays) or $\Delta C = 2$ (mixing) 
transitions. In the case of $\DzDzb$ mixing these operators generate contributions 
to the effective operators that change $D^0$ state into $\Dzb$ state,
leading to the mass eigenstates
\begin{equation} \label{definition1}
| D_{1,2} \rangle =
p | D^0 \rangle \pm q | \bar D^0 \rangle,
\end{equation}
where the complex parameters $p$ and $q$ are obtained from diagonalizing 
the $D^0-\Dzb$ mass matrix. Note that $|p|^2 + |q|^2 = 1$. If CP-violation
in mixing is neglected, $p$ becomes equal to $q$, so $| D_{1,2} \rangle$ 
become $CP$ eigenstates, $CP | D_{\pm} \rangle = \pm | D_{\pm} \rangle$.
The mass and width splittings between these eigenstates are given by
\begin{eqnarray} \label{definition}
x \equiv \frac{m_2-m_1}{\Gamma}, ~~
y \equiv \frac{\Gamma_2 - \Gamma_1}{2 \Gamma}.
\end{eqnarray}
It is known experimentally that $\DzDzb$ mixing proceeds extremely slowly, 
which in the Standard Model is usually attributed to the absence of superheavy 
quarks destroying GIM cancellations.

\section{Phenomenology of $\DzDzb$ mixing}

{\bf Semileptonic decays.} The most natural way to search for charm mixing is to employ 
semileptonic decays. It is also not the most sensivite way, as it is only sensitive to 
$R_D=(x^2+y^2)/2$, a quadratic function of $x$ and $y$. Use of the $D^0$ semileptonic decays for the mixing 
search involves the measurement of the time-dependent or time-integrated rate for the 
wrong-sign (WS) decays of $D$, where $c\to(\overline{c}\mbox{ via mixing})\to\overline{s}\ell^-\overline{\nu}$, 
relative to the right-sign (RS) decay rate, $c\to s\ell^+\nu$. 

{\bf Nonleptonic decays to non-CP eigenstates.} Currently the most stringent 
limits on the $D^0$ mixing parameters arise from the decay time dependent 
$D^0\to K^+\pi^-$ measurements. Time-dependent studies allow separation 
of the direct doubly-Cabbibo suppressed (DCS) $D^0\to K^+\pi^-$ amplitude from 
the mixing contribution $D^0 \to \Dzb \to K^+ \pi^-$,
\begin{eqnarray}\label{Kpi}
\Gamma_{D^0 \to K^+ \pi^-}
=e^{-\Gamma t}|A_{K^-\pi^+}|^2 
~\left[
R+\sqrt{R}R_m(y'\cos\phi-x'\sin\phi)\Gamma t
+R_m^2 R_D^2 (\Gamma t)^2
\right],
\end{eqnarray}
where $R$ is the ratio of DCS and Cabibbo favored (CF) decay rates. 
Since $x$ and $y$ are small, the best constraint comes from the linear terms 
in $t$ that are also {\it linear} in $x$ and $y$.
A direct extraction of $x$ and $y$ from Eq.~(\ref{Kpi}) is not possible due 
to unknown relative strong phase $\delta_D$ of DCS and CF 
amplitudes, as $x'=x\cos\delta_D+y\sin\delta_D$, $y'=y\cos\delta_D-x\sin\delta_D$. 
This phase can be measured independently. The corresponding formula can 
also be written\cite{Bergmann:2000id} for $\Dzb$ decay with $x' \to -x'$ and 
$R_m \to R_m^{-1}$.

{\bf Nonleptonic decays to CP eigenstates.} $D^0$ mixing can be measured by comparing 
the lifetimes extracted from the analysis of $D$ decays into the CP-even and CP-odd 
final states. In practice, lifetime measured in $D$ decays into CP-even final state $f_{CP}$,  
such as $K^+K^-, \pi^+\pi^-, \phi K_S$, etc., is compared to the one obtained from a 
measurement of decays to a non-CP eigenstate, such as $K^-\pi^+$. This analysis is also 
sensitive to a {\it linear} function of $y$ via
\begin{equation}
y_{CP} = \frac{\tau(D \to K^-\pi^+)}{\tau(D \to K^+K^-)}-1=
y \cos \phi - x \sin \phi \left[\frac{R_m^2-1}{2}\right],
\end{equation}
where $\phi$ is a CP-violating phase. In the limit of vanishing CP violation 
$y_{CP} = y$. This measurement requires precise determination of lifetimes.

{\bf Quantum-correlated final states.} The construction of tau-charm factories introduces 
new {\it time-independent} methods that are sensitive to a linear function of $y$. One 
can use the fact that heavy meson pairs produced in the decays of heavy quarkonium 
resonances have the useful property that the two mesons are in the CP-correlated 
states\cite{Atwood:2002ak}. For instance, by tagging one of the mesons as a CP 
eigenstate, a lifetime difference may be determined by measuring the leptonic 
branching ratio of the other meson. The final states reachable by neutral charmed 
mesons are determined by a set of selection rules according to the initial virtual 
photon quantum numbers $J^{PC}=1^{--}$\cite{Atwood:2002ak}. 
Since we know whether this $D(k_2)$ state is 
tagged as a (CP-eigenstate) $D_\pm$ from the decay of $D(k_1)$ to a 
final state $S_\sigma$ of definite CP-parity $\sigma=\pm$, we can 
easily determine $y$ in terms of the semileptonic branching ratios of $D_\pm$, which
we denote $\cal{B}^{\ell}_\pm$. Neglecting small CP-violating effects,
\begin{equation} 
y=\frac{1}{4} \left(
\frac{\cal{B}^{\ell}_+(D)}{\cal{B}^{\ell}_-(D)}-
\frac{\cal{B}^{\ell}_-(D)}{\cal{B}^{\ell}_+(D)}
\right).
\end{equation}
A more sophisticated version of this formula as well as studies of 
feasibility of this method can be found in Ref.~[\refcite{Atwood:2002ak}].

The current experimental bounds on $y$ and $x$ are~\cite{pdg}
\begin{eqnarray*}
y  < 0.008 \pm 0.005 \ , \quad \quad
 x < 0.029 \ \
(95\%~ {\rm C.L.}) \nonumber\ \ .
\end{eqnarray*}
%

\section{Charm mixing predictions in the Standard Model}

The current experimental upper bounds on $x$ and $y$ are on the order of 
a few times $10^{-3}$, and are expected to improve in the coming
years.  To regard a future discovery of nonzero $x$ or $y$ as a signal for new 
physics, we would need high confidence that the Standard Model predictions lie
well below the present limits.  As was shown in [\refcite{Falk:2001hx}], 
in the Standard Model, $x$ and $y$ are generated only at second order in SU(3)$_F$ 
breaking, 
\begin{equation}
x\,,\, y \sim \sin^2\theta_C \times [SU(3) \mbox{ breaking}]^2\,,
\end{equation}
where $\theta_C$ is the Cabibbo angle.  Therefore, predicting the
Standard Model values of $x$ and $y$ depends crucially on estimating the 
size of SU(3)$_F$ breaking. 

%
\begin{figure}[tb]
\begin{center}
\includegraphics[scale=0.32]{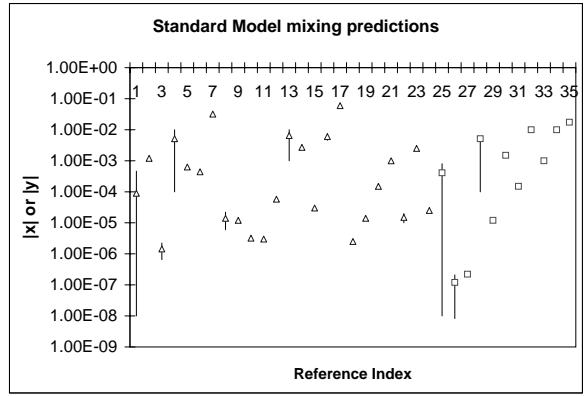}
\caption{Standard Model predictions for $|x|$ (open triangles) 
and $|y|$ (open squares).}
\label{fig:DDmixSM}
\end{center}
\end{figure}
Theoretical predictions of $x$ and $y$ within
the Standard Model span several orders of magnitude\footnote{Horizontal line references are 
tabulated in Table 5 of Ref.~[\refcite{Predictions}].} (see Fig.~\ref{fig:DDmixSM}).
Roughly, there are two approaches, neither of which give very reliable
results because $m_c$ is in some sense intermediate between heavy and
light.  The ``inclusive'' approach is based on the operator
product expansion (OPE).  In the $m_c \gg \Lambda$ limit, where
$\Lambda$ is a scale characteristic of the strong interactions, $\Delta
M$ and $\Delta\Gamma$ can be expanded in terms of matrix elements of local
operators\cite{Inclusive}.  Such calculations typically yield $x,y < 10^{-3}$.  
The use of the OPE relies on local quark-hadron duality, 
and on $\Lambda/m_c$ being small enough to allow a truncation of the series
after the first few terms.  The charm mass may not be large enough for these 
to be good approximations, especially for nonleptonic $D$ decays.
An observation of $y$ of order $10^{-2}$ could be ascribed to a
breakdown of the OPE or of duality,  but such a large
value of $y$ is certainly not a generic prediction of OPE analyses.
The ``exclusive'' approach sums over intermediate hadronic
states, which may be modeled or fit to experimental data\cite{Exclusive}.
Since there are cancellations between states within a given $SU(3)$
multiplet, one needs to know the contribution of each state with high 
precision. However, the $D$ is not light enough that its decays are dominated
by a few final states.  In the absence of sufficiently precise data on many decay 
rates and on strong phases, one is forced to use some assumptions. While most 
studies find $x,y < 10^{-3}$, Refs.~[\refcite{Exclusive}] obtain $x$ and 
$y$ at the $10^{-2}$ level by arguing that SU(3)$_F$ violation is of order
unity. 
It was also shown that phase space effects alone provide enough SU(3)$_F$ 
violation to induce $x,y\sim10^{-2}$~[\refcite{Falk:2001hx}].
Large effects in $y$ appear for decays close to $D$ threshold, where
an analytic expansion in SU(3)$_F$ violation is no longer possible; a dispersion 
relation can then be used to show that $x$ would receive contributions of
similar order of magnitude.

The above discussion shows that, contrary to $B$ and $K$ systems, theoretical 
calculations of $x$ and $y$ are quite uncertain\cite{Reviews}, and the values
near the current experimental bounds cannot be ruled out. 

\section{New Physics contribution to $\DzDzb$ mixing}

As one can see from the previous discussion, mixing in the charm system is very small.
As it turns out, theoretical predictions of $x$ and $y$ in the Standard Model
are very uncertain, from a percent to orders of magnitude smaller\cite{Falk:2001hx,Predictions}.
Thus, New Phyiscs (NP) contributions can easily stand out.

In order to see how NP might affect the mixing amplitude, it is instructive to
consider off-diagonal terms in the neutral D mass matrix,
\begin{eqnarray}\label{M12}
2 M_{\rm D} \left (M - \frac{i}{2}\, \Gamma\right)_{12} =
  \langle \D0bar | 
{\cal H}_w^{\Delta C=-2} | D^0 \rangle 
+ \sum_n {\langle \D0bar | {\cal H}_w^{\Delta
  C=-1} | n \rangle\, \langle n | {\cal H}_w^{\Delta C=-1} 
| D^0 \rangle \over M_{\rm D}-E_n+i\epsilon}\,
\end{eqnarray}
where ${\cal H}_w^{\Delta C=-1}$ is the effective $|\Delta C| = 1$ hamiltonian. 

{\bf New Phyiscs in $|\Delta C|=2$ interactions.} Since
all new physics particles are much heavier than the Standard Model ones, the most 
natural place for NP to affect mixing amplitudes is in the $|\Delta C|=2$ piece, which 
corresponds to a local interaction at the charm quark mass scale. Integrating out NP degrees
of freedom at some scale $\Lambda$, we are left with an effective Hamiltonian written 
in the form of series of operators of increasing dimension\cite{NewPaper}. 
Realizing this, it is not hard to write the complete basis of those effective operators,
which most conveniently can be done in terms of left- and right-handed quark fields,
\beq\label{SeriesOfOperators}
{\cal H}^{\Delta C=2}_{NP} = \sum_{i=1} C_i(\mu) ~{\cal Q}_i(\mu),
\eeq
where $C_i$ are the Wilson coefficients, and $Q_i$ are the effective operators,
\bea\label{SetOfOperators}
{\cal Q}_1 &=& \overline{u}_L \gamma_\mu c_L \overline{u}_L \gamma^\mu c_L, \qquad
{\cal Q}_5 = \overline{u}_R  \sigma_{\mu\nu} c_L \overline{u}_R \sigma^{\mu\nu} c_L
\nonumber \\
{\cal Q}_2 &=& \overline{u}_R \gamma_\mu c_R \overline{u}_L \gamma^\mu c_L, \qquad
{\cal Q}_6 = \overline{u}_R \gamma_\mu c_R \overline{u}_R \gamma^\mu c_R,
\\
{\cal Q}_3 &=& \overline{u}_L c_R \overline{u}_R c_L, \qquad\qquad
{\cal Q}_7 = \overline{u}_L c_R \overline{u}_L c_R,
\nonumber \\
{\cal Q}_4 &=& \overline{u}_R c_L \overline{u}_R c_L, \qquad\qquad
{\cal Q}_8 = \overline{u}_L  \sigma_{\mu\nu} c_R \overline{u}_L \sigma^{\mu\nu} c_R,
\nonumber 
\eea
These operators exhaust the list of possible contributions to $\Delta C=2$ transitions.
Since these operators are generated at the scale $\mu = \Lambda$ (at which new physics is
integrated out), a non-trivial operator mixing can occur if we take into account
renormalization group running of these operators between $\mu=\Lambda$ and
$\mu\simeq m_c$ scales. This running can be accounted for by solving RG equations
obeyed by the Wilson coefficient functions,
\beq\label{RGE}
\frac{d}{d \log \mu} \vec C (\mu) = \hat \gamma^T(\mu) \vec C (\mu),
\eeq
where $\hat \gamma^T(\mu)$ represents the matrix of anomalous dimensions of
operators of  Eq.~(\ref{SetOfOperators}).
A prediction for a mixing parameter $x$ in a particular model of new physics is then 
obtained by computing $C_i (\Lambda)$ for a set of ${\cal Q}_i (\Lambda)$ generated by 
a given model, running the RG equations of Eq.(~\ref{RGE}) and computing matrix elements 
$\langle \Dzb | {\cal Q}_i (m_c)| \Dz \rangle$.
\begin{figure}[tb]
\begin{center}
\includegraphics[scale=0.32]{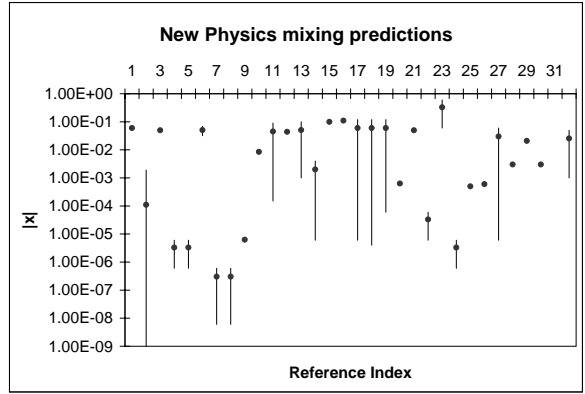}
\caption{New Physics predictions for $|x|$.}
\label{fig_NP}
\end{center}
\end{figure}
As can be seen from Fig.~(\ref{fig_NP})\footnote{Horizontal line references are 
tabulated in Table 5 of Ref.~[\refcite{Predictions}].}, predictions for $x$ vary by orders of 
magnitude for 
different models. It is interesting to note that some models {\it require} large signals 
in the charm system if mixing and FCNCs in the strange and beauty systems are to be small 
(e.g. the SUSY alignment model). 

{\bf New Phyiscs in $|\Delta C|=1$ interactions.}  The local $|\Delta C|=2$ interaction 
cannot, however, affect $\Delta \Gamma_{\rm D}$ because 
it does not have an absorptive part. Thus, naively, NP cannot affect lifetime difference $y$.
This is, however, not quite correct. Consider a $D^0$ decay amplitude which includes
a small NP contribution, $A[D^0 \to n]=A_n^{\rm (SM)} + A_n^{\rm (NP)}$.
Here, $A_n^{\rm (NP)}$ is assumed to be smaller than the current experimental
uncertainties on those decay rates. Then it is a good approximation to write $y$ as
\begin{eqnarray}\label{schematic}
y &\simeq& \sum_n \frac{\rho_n}{\Gamma_{\rm D}} 
A_n^{\rm (SM)} \bar A_n^{\rm (SM)}
+ 2\sum_n \frac{\rho_n}{\Gamma_{\rm D}}
A_n^{\rm (NP)} \bar A_n^{\rm (SM)} \ \ . 
\label{approx}
\end{eqnarray}
The SM contribution to $y$ is known to vanish in the limit of exact flavor $SU(3)$.
Moreover, the first order correction is also absent, so the SM contribution arises 
only as a {\it second} order effect. Thus, those NP contributions which do not vanish 
in the flavor $SU(3)$ limit must determine the lifetime difference there,
even if their contibutions are tiny in the individual decay amplitudes\cite{Golowich:2006gq}. 
A simple calculation reveals that NP contribution to $y$ can be as large as several percent
in R-parity-violating SUSY models or as small as $\sim 10^{-10}$ in the models with interactions 
mediated by charged Higgs particles\cite{Golowich:2006gq}.

Consider, for example, models that do not vanish in the $SU(3)$ limit. The two most common 
scenarios involve models whose low energy $|\Delta C|=1$ effective hamiltonian involves
(V-A)$\otimes$(V-A) and (S-P)$\otimes$(S+P) couplings. For instance, SUSY models without 
R-parity\cite{bt} would fit the bill. In these models, there are flavor-changing interactions 
of sleptons that can be obtained from the lagrangian
\bea
{\cal L}_{\not{R}} \ = \  \lambda_{ijk}'L_i Q_j D^c_k \ \ ,
\eea
The slepton-mediated interaction is not suppressed in the flavor SU(3) 
limit and leads to 
\bea\label{RPVSUSY}
y_{\not{R}}=
\frac{C' \widetilde{\lambda}}{M_{\widetilde\ell}^2}
\left[
\left( C_2 - 2 C_1\right) \langle Q^\prime\rangle +
\left( C_1 - 2 C_2\right) \langle \widetilde Q^\prime\rangle
\right],
\eea
where $C'=-G_F m_c^2/(6 \sqrt{2} \pi M_{\rm D} \Gamma_{\rm D})$, 
$M_{\widetilde\ell}$ is a slepton mass,  
$\widetilde\lambda$ is given by $\widetilde{\lambda}=
\lambda_{sd} - \lambda \left(\lambda_{dd}-\lambda_{ss}\right)-
\lambda^2 \left(\lambda_{ds} + \lambda_{sd}\right)$
with $\lambda_{sd}=\lambda'_{i12}\lambda'_{i21} \le 1 \times 10^{-9}$, 
$\lambda_{ss}=\lambda'_{i11}\lambda'_{i21} \le 5 \times 10^{-5}$,
$\lambda_{dd}=\lambda'_{i21}\lambda'_{i22} \le 5 \times 10^{-5}$, 
$\lambda_{ds}=\lambda'_{i11}\lambda'_{i22} \le 5 \times 10^{-2}$ 
(see [\refcite{bt}]), and $\langle Q^\prime\rangle$ is
\bea\label{Qprime}
\langle Q' \rangle  \ &=& \ \langle \overline D^0 |  \overline{u}_{i} 
\gamma_\mu P_L c_{i}
\ \overline{u}_{j} \gamma^\mu P_R c_{j} | D^0 \rangle \ \ .
\eea
Operators with a tilde are obtained by swapping color indices in the 
charm quark operators. Using factorization to estimate matrix elements 
of the above operators and taking for definiteness 
$M_{\widetilde\ell} = 100$~GeV, we arrive at
$y_{\not{R}} \simeq -3.7\%$\cite{Golowich:2006gq}. 

On the other hand, there are also several reasons that some NP models 
vanish in the flavor $SU(3)$ limit.  First, the structure of the
NP interaction might simply mimic the one of the SM. Effects like that 
can occur in some models with extra space dimensions. Second,
the chiral structure of a low-energy effective lagrangian in 
a particular NP model could be such that the leading, 
mass-independent contribution vanishes exactly, 
as in a left-right model (LRM). Third, the NP coupling might explicitly
depend on the quark mass, as in a model with multiple
Higgs doublets. However, most of these models feature 
second order $SU(3)$-breaking already at leading order in the $1/m_c$ 
expansion. This should be contrasted with the SM, where the 
leading order is suppressed by six powers of $m_s$ and the second order 
only appears as a $1/m_c^6$-order correction.

For instance, LRM provide new tree-level contributions mediated by right-handed ($W^{\rm (R)}$) 
bosons\cite{mp}. The relevant effective lagrangian is
\beq
{\cal L}_{\rm LR}=-\frac{g_{\rm R}}{\sqrt{2}} \ {\bf V}_{ab}^{\rm (R)} \
\overline{u}_{a,i} \gamma^\mu P_R d_{b,i}~ W^{\rm (R)}_\mu +
\mbox{h.c.}  \ \ , 
\eeq
where ${\bf V}^{\rm (R)}_{ik}$ are the coefficients of the 
right-handed CKM matrix.
Since current experimental limits allow $W^{\rm (R)}$ masses as low as 
a TeV\cite{pdg}, a sizable contribution to $y$ is quite possible. We obtain 
\bea
& & y_{\rm LR} = - C_{\rm LR} {\bf V}^{\rm (R)}_{cs} 
{\bf V}^{\rm (R)*}_{us} 
\left[ {\cal C}_1 \langle Q' \rangle
+ {\cal C}_2 \langle \tilde Q' \rangle 
\right] \ \ , \label{yLRM}
\eea
where $C_{\rm LR} \equiv \lambda G_F^{\rm (R)} 
G_F m_c^2 x_s/(\pi M_{\rm D}\Gamma_{\rm D})$, 
$G_F^{\rm (R)}/\sqrt{2} \equiv g^2_{\rm R}/8 M^2_{W_{\rm R}}$, 
${\cal C}_{1,2}$ are the SM Wilson coefficients and the operators
appearing in Eq.~(\ref{yLRM}) are given in Eq.~(\ref{Qprime}).
Using [\refcite{pdg}], we obtain numerical values for two possible 
realizations: (i) 'Manifest LR' (${\bf V}^{\rm (L)} = {\bf V}^{\rm (R)}$) 
gives $y_{\rm LR} = -4.8 \cdot 10^{-6}$ with $M_{W_{\rm R}} = 1.6$~TeV 
and (ii) 'Nonmanifest LR' (${\bf V}^{\rm (R)}_{ij} \sim 1$) gives 
$y_{\rm LR} = -8.8 \cdot 10^{-5}$ with $M_{W_{\rm R}} = 0.8$~TeV\cite{Golowich:2006gq}.

As one can wee, small NP contributions to $|\Delta C|=1$ processes 
produce substantial effects in the $D^0\overline{D}^0$ lifetime difference, 
even if such contributions are currently undetectable in the experimental 
analyses of charmed meson decays. Coupled with a known difficulty in 
computing SM contributions to D-meson decay amplitudes, it might be 
advantageous to use experimental constraints on $y$ in order to test 
various NP scenarios due to better theoretical control over the NP contribution 
and SU(3) suppression of the SM amplitude.

{\bf New Phyiscs and CP-violation in $D^0\overline{D}^0$ mixing.}  Another 
possible manifestation of new physics interactions in the charm
system is associated with the observation of (large) CP-violation. This 
is due to the fact that all quarks that build up the hadronic states in weak 
decays of charm mesons belong to the first two generations. Since $2\times2$ 
Cabbibo quark mixing matrix is real, no CP-violation is possible in the
dominant tree-level diagrams which describe the decay amplitudes. 
CP-violating amplitudes can be introduced in the Standard Model by including 
penguin or box operators induced by virtual $b$-quarks. However, their 
contributions are strongly suppressed by the small combination of 
CKM matrix elements $V_{cb}V^*_{ub}$. It is thus widely believed that the 
observation of (large) CP violation in charm decays or mixing would be an 
unambiguous sign for new physics\cite{Predictions,Petrov:2004gs}. 

\section{Summary}

Charm physics provides new and unique opportunities for indirect searches for 
New Physics. NP can affect charm mixing in a variety of ways, mainly
affecting both $x$ and $y$, as well as providing CP-violating asymmetries.
Expected large statistical samples of charm data will allow new sensitive
measurements of charm mixing and CP-violating parameters.

\section*{Acknowledgments}

This work was supported in part by the U.S.\ National Science Foundation
CAREER Award PHY--0547794, and by the U.S.\ Department of Energy under Contract
DE-FG02-96ER41005.



\begin{thebibliography}{99}


\bibitem{Petrov:1997ch}
A.~Datta, D.~Kumbhakar,
Z.\ Phys.\ C{\bf 27}, 515 (1985);
A.~A.~Petrov,
Phys.\ Rev.\ D{\bf 56}, 1685 (1997);
E.~Golowich and A.~A.~Petrov,
Phys.\ Lett.\ B {\bf 427}, 172 (1998);
  E.~Golowich and A.~A.~Petrov,
  Phys.\ Lett.\ B {\bf 625}, 53 (2005).

\bibitem{Bergmann:2000id}
S.~Bergmann, {\it et. al},
Phys.\ Lett.\ B {\bf 486}, 418 (2000);
A.~F.~Falk, Y.~Nir and A.~A.~Petrov,
JHEP {\bf 9912}, 019 (1999).

\bibitem{Atwood:2002ak}
D.~Atwood and A.~A.~Petrov,
Phys.\ Rev.\ D {\bf 71}, 054032 (2005);
M.~Gronau, Y.~Grossman and J.~L.~Rosner,
Phys.\ Lett.\ B {\bf 508}, 37 (2001);
D.~M.~Asner and W.~M.~Sun,
Phys.\ Rev.\ D {\bf 73}, 034024 (2006)


\bibitem{pdg} 
  W.~M.~Yao {\it et al.}  [Particle Data Group],
  J.\ Phys.\ G {\bf 33} (2006) 1.

\bibitem{Falk:2001hx}
A.~F.~Falk, Y.~Grossman, Z.~Ligeti and A.~A.~Petrov,
Phys.\ Rev.\ D {\bf 65}, 054034 (2002);
A.~F.~Falk, {\it et. al},
Phys.\ Rev.\ D {\bf 69}, 114021 (2004).

\bibitem{Predictions}
A.~A.~Petrov,
eConf {\bf C030603}, MEC05 (2003)
[arXiv:hep-ph/0311371];
H.~N.~Nelson,
in {\it Proc. of the 19th Intl. Symp. on Photon and Lepton 
Interactions at High Energy LP99 } ed. J.A. Jaros and M.E. Peskin,
arXiv:hep-ex/9908021.

\bibitem{Inclusive}
H.~Georgi, Phys. Lett. B297, 353 (1992);
T.~Ohl, G.~Ricciardi and E.~Simmons, Nucl. Phys. B403, 605 (1993);
I.~Bigi and N.~Uraltsev,
Nucl.\ Phys.\ B {\bf 592}, 92 (2001).

\bibitem{Exclusive}
J. Donoghue, E. Golowich, B. Holstein and J. Trampetic,
Phys. Rev. D33, 179 (1986);
L. Wolfenstein, Phys.\ Lett.\ B164, 170 (1985);
P. Colangelo, G. Nardulli and N. Paver,  Phys.\ Lett.\ B242, 71 (1990);
T.A. Kaeding,  Phys. Lett. B357, 151 (1995);
A.~A.~Anselm and Y.~I.~Azimov,
Phys.\ Lett.\ B {\bf 85}, 72 (1979).

\bibitem{Reviews}
S.~Bianco, F.~L.~Fabbri, D.~Benson and I.~Bigi,
Riv.\ Nuovo Cim.\  {\bf 26N7-8}, 1 (2003);
G.~Burdman and I.~Shipsey,
Ann.\ Rev.\ Nucl.\ Part.\ Sci.\  {\bf 53}, 431 (2003).

\bibitem{Petrov:2004gs}
A.~A.~Petrov,
Phys.\ Rev.\ D {\bf 69}, 111901 (2004).

\bibitem{NewPaper}
  J.~Hewett, E.~Golowich, S.~Pakvasa and A.~A.~Petrov, to be published.

\bibitem{Golowich:2006gq}
  E.~Golowich, S.~Pakvasa and A.~A.~Petrov,
  arXiv:hep-ph/0610039.


\bibitem{bt} {\it E.g.}, see H. Baer and X.Tata, {\it Weak Scale 
  Supersymmetry: From Superfields to Scattering Events} 
  (Cambridge University Press, Cambridge, England 2006). 	
	
\bibitem{mp}
  R.~Mohapatra and J.~Pati,
  Phys.\ Rev.\ D {\bf 11}, 566 (1975).
  


\end{thebibliography}
\end{document}